# Harnessing Short-Range Surface Plasmons in Planar Silver Films via Disorder-Engineered Metasurfaces

*Maximilian Buchmüller, Ivan Shutsko, Sven Oliver Schumacher and Patrick Görrn\**

Maximilian Buchmüller, Ivan Shutsko, Sven Oliver Schumacher, Patrick Görrn
Chair of Large Area Optoelectronics, University of Wuppertal, Rainer-Gruenter-Str. 21, 42119 Wuppertal, Germany
\*E-mail: goerrn@uni-wuppertal.de

Maximilian Buchmüller, Ivan Shutsko, Sven Oliver Schumacher, Patrick Görrn
Wuppertal Center for Smart Materials and Systems, University of Wuppertal, Rainer-Gruenter-Str. 21, 42119 Wuppertal, Germany

**Abstract**

Short-range surface plasmon polaritons (SR-SPPs) can arise due to the hybridization of surface plasmon polaritons propagating along the two interfaces of a thin metal slab. In optics, they have gained particular interest for imaging and sensing applications, because of their short wavelengths at optical frequencies along with strong field enhancement. However, mediating the interaction of SR-SPPs with photons in planar films is difficult because of the large momentum mismatch. For efficient coupling, nanostructuring such thin films (~20nm thickness), or placing metallic nanostructures in close proximity to the planar film is technologically challenging and can strongly influence the SR-SPP properties. In this paper, harnessing SR-SPPs in planar silver films is demonstrated using disorder-engineered metasurfaces. The disorder-engineering is realized by the light-controlled growth of silver nanoparticles. The dispersion of the hybrid modes with the silver thickness is measured and compared with simulations. We anticipate these results to introduce a novel and facile method for harnessing SR-SPPs in planar optical systems and make use of their promising properties for imaging, sensing and nonlinear optics.

1. Introduction

Surface plasmon polaritons (SPPs) are electromagnetic surface waves propagating along a metal/dielectric interface, which can be excited by the coupling of photons to collective oscillations of free electrons in the metal.[1] Perpendicular to the interface, SPPs are confined evanescently on sub-wavelength length scales, while their propagation length is governed by radiation loss and absorption loss inside the metal.[2] Due to their strong field confinement beyond the optical diffraction limit, SPPs enable immense electromagnetic field enhancement inside nanoscale mode volumes.[3] As a result, SPPs have gained remarkable attention in modern optics with a wide range of applications, including surface enhanced Raman spectroscopy,[4,5] nonlinear optics,[6] and sensing.[7]

In many of these applications, noble metal (e.g. silver) films are used together with other functional layers in a planar stack geometry. Such planar metal films support two surface plasmon modes, one at each interface of the film. For large film thicknesses of the metal $t_m$, the two SPPs can be treated as individual single-interface surface plasmons with wavevectors governed by the respective permittivity profiles adjacent to each interface. If the metal film becomes sufficiently thin ($t_m$ in the order of a few tens of nanometers), the two SPP modes can hybridize. This hybridization can be described by a coupled system with two eigenmodes, showing dispersion with the film thickness $t_m$.[8] While one eigenmode extends far into the dielectric, the other eigenmode is strongly confinement to the metal slab, and thus possesses significantly higher absorption loss. Therefore, the two eigenmodes are often referred to as *long-range* (LR) and *short-range* (SR) surface plasmons, because of their comparatively long and short propagation lengths, respectively.

Given their strong field enhancement and short wavelengths, SR-SPPs have been used for sensing,[9] focusing,[10,11] and high-resolution microscopy.[12] The formation of hybrid bound states in the continuum (BICs) has also been investigated recently, when SR-SPPs are coupled to photonic waveguide modes.[13] Although being of high interest, SR-SPPs have received less attention compared to LR-SPPs in the past.[14] On the one hand, SR-SPPs are relatively difficult to address with photons from free-space. Compared to photons, surface plasmons exhibit larger momentum for a given frequency. Thus, a photon incident on a metal surface requires additional lateral momentum for being coupled to a surface plasmon. While for single interface SPPs, this additional momentum can be provided by a high index prism (Kretschmann-Raether geometry), the exploitation of SR-SPPs in ultrathin planar silver films requires surface nanostructures (e.g. gratings), because of the high effective index.[2] The effective index even strongly increases with decreasing $t_m$.[15] At the same time due to their large field confinement SR-SPPs are extremely sensitive to imperfections like surface roughness.[16] Therefore, the fabrication of nanostructures required for efficient coupling to SR-SPPs becomes even more complex and technologically challenging.

In the scope of sensing, both a spectrally narrow plasmon resonance (high $Q$-factor) as well as a large spectral shift upon a change in refractive index (sensitivity $S$) are desired simultaneously to reach satisfying sensing performance. Although SR-SPPs have the opportunity for high sensitivities due to the strong field enhancement, they are spectrally broad due to the short propagation lengths caused by absorption. The nanostructuring of the ultra-thin silver film for coupling even causes additional radiation loss leading to an even broader plasmon resonance.[2]

Recently it has been shown that a disorder-engineered metasurface consisting of silver nanoparticles (AgNPs) in close proximity to a thin silver film can enable high optical sensing performance.[17] Due to the selective scattering mediated by the metasurface, the SPP modes propagating on the silver film create a sharp resonance in the reflection signal, which is highly sensitive to the refractive index of the analyte. The width of this resonance is around two orders of magnitude narrower compared to the initial SPP resonance without the metasurface and is largely governed by the divergence of the laser used for fabrication and probing. It can be concluded from these results that disorder-engineered metasurfaces can enable to generate high-$Q$ resonances in plasmonic structures, which still exploit the high sensitivity of SPPs resulting in high sensing performance. Following this thought, the ability of addressing SR-SPPs with their enormous field confinement and sensitivity using such metasurfaces could be an important step towards the detection of single binding events.

In this paper, we demonstrate a method to exploit SR-SPPs in smooth ultra-thin silver films using a Kretschmann-Raether geometry and a disorder-engineered metasurface providing additional momentum. The metasurfaces are fabricated by the plasmon-mediated growth of AgNPs from solution and are inherently adapted to their electromagnetic environment. From the morphology of the metasurface, we show that LR-SPPs and SR-SPPs can be addressed by using a laser beam at a fixed incident angle. Further, we analyse the dispersion of both modes with respect to the silver thickness $t_m$ in order to confirm the largely undisturbed propagation of SR-SPPs in the system, which we consider highly promising to increase selectivity and performance in low-cost optical sensors.

## 2. Results and Discussion

The experimental setup is shown in **Figure 1a**. A thin silver film ($t_m = 56$ nm) on a sapphire substrate is mounted onto a sapphire prism. A p-polarized laser is then aligned with respect to the SPP phase matching conditions. We find the maximum excitation efficiency of SPPs (minimum reflection of the laser) under an incident angle of $\Theta_{in} = 54°$ for a laser wavelength of $\lambda_0 = 660$ nm. AgNPs are then grown from solution by electro-less deposition (ELD) of silver, as reported previously.[17,18] The resulting plasmonic metasurface is analyzed by scanning electron microscopy (SEM). A typical scanning electron micrograph is shown in Figure 1b. From the SEM investigations, the average particle radius is found to be 87 nm. Considering the real space, the metasurface appears to contain no deterministic order. However, features of engineered disorder, e.g. disordered hyperuniformity, have recently been found in metasurfaces fabricated with the method presented here.[19] A well-established method to investigate the morphology of metasurfaces is given by the Fourier-transform[20],[21]. In Figure 1c, a typical Fourier-transformed electron micrograph (FTEM) of the metasurface is shown. Two pronounced dark rings are visible that touch each other at $x^{-1} = y^{-1} = 0$, termed *structure rings*. As we have shown in previous work, the symmetric shift of the two rings along the $x^{-1}$-axis $r_{in}$ as well as the radius $r_s$ of the rings (emphasized by white arrows in Figure 1c) can be attributed to the surface plasmon polariton wave involved in the metasurface growth.[17,22] While $r_{in}$ is determined by the incident plane wave, $r_s$ represents the spherical waves scattered by the metasurface. Here, both $r_{in}$ and $r_s$ can be attributed to the SPP$_a$ mode propagating along the silver/PMMA interface as $r_{in} = r_s = 2.14 \ \mu m^{-1} = 1/\lambda_{SPPa}$. The corresponding effective index of the SPP$_a$ mode is given by

$n_{SPPa} = \lambda_0/\lambda_{SPPa} = 1.41$. We do not observe any features of the SPP$_b$ mode. Apparently, it does not interact with the metasurface because of the large silver film thickness.

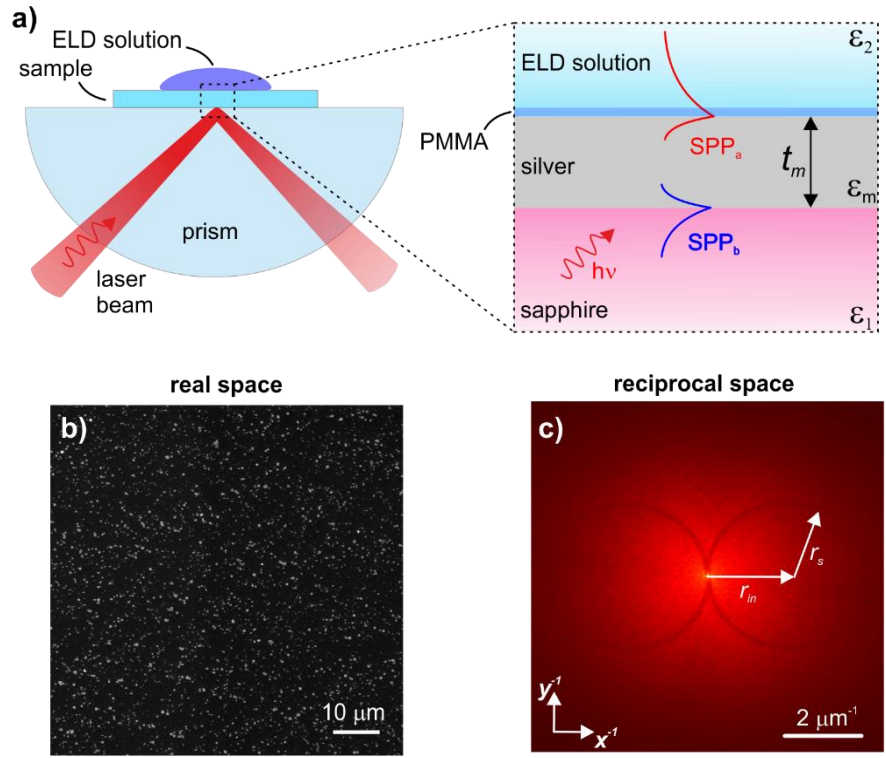

Figure 1: Experimental setup and results of SEM investigations of disorder-engineered metasurfaces fabricated on thick silver films ($t_m = 56$ nm). a) Sketch of the Kretschmann-Raether setup used for the experiments and the stack geometry of the samples (inset) emphasising the two plasmon modes propagating at both interfaces of the silver film (blue and red); b) A typical SEM image of the fabricated metasurfaces.; c) Fourier-transform of typical SEM images revealing the engineered disorder of the metasurfaces. The white arrows indicate the horizontal shift $r_{in}$ and the radius $r_s$ of the structure rings.

In the next step, we decrease the silver film thickness to $t_m = 25$ nm for enabling plasmon hybridization. The resulting layer stack together with qualitative one-dimensional field profiles of the LR (red) and SR (blue) SPP modes are shown in **Figure 2a**. In addition, we simulated the two-dimensional field distribution of the SR-SPP mode using rigorous coupled wave-analysis (RCWA). In the RCWA, we used a refractive index for the ELD solution of $n_{ELD} = 1.336$, which has been measured using Michelson-interferometry in an earlier work.[17] The simulation results are shown in Figure 2b and 2c displaying the magnetic field distribution and the magnetic field intensity. It becomes visible that the SR-SPP mode is predominantly bound to the silver/sapphire interface (see Figure 2b). However, both modes (LR- and SR-SPP) extend into the ELD solution and can thus interact with the later metasurface. Thus, after the growth of that metasurface also both modes can interact with an analyte placed on top of the device. From the RCWA, we obtained a plasmon wavelength of $\lambda_{SR} = \frac{660 \text{ nm}}{2.13} = 310$ nm for $t_m = 25$ nm with a corresponding penetration depth (1/e intensity) into the ELD solution (normal to the interface) of 31 nm with respect to the PMMA/ELD interface. The small penetration depth of the SR SPP into the ELD solution shows the strong field confinement to the metal slab indicating its promising properties to achieve high surface sensitivities. Thus, the SR-SPPs would be influenced predominantly by surface binding events and the LR-SPPs would rather detect bulk permittivity fluctuations. In summary, in the simulation we find two modes with different effective indices that can interact with the growing metasurface.

In experiment, the reflection of the laser exhibits a minimum at an incident angle of $\theta_{in} = 54°$ for $\lambda_0 = 660$ nm and the metasurface is prepared in the same way as reported for thick silver films. In the FTEM, we now observe four instead of two structure rings (see Figure 2d). The inner structure rings again touch each other at $x^{-1} = y^{-1} = 0$. In this case, we found that both the radii and the horizontal shift of the inner structure rings correspond to the LR-SPP mode with $r_{in} = r_{s,1} = 2.16$ µm$^{-1} = 1/\lambda_{LR}$ and an effective index of 1.42. The radii of the outer structure rings are larger than the horizontal shift ($r_{s,2} > r_{in}$) und thus these structure rings possess two intersections on the $x^{-1} = 0$ axis. We found that the outer structure rings correspond to the SR-SPP mode as $r_{s,2} = 3.18\ \mu m^{-1} = 1/\lambda_{SR}$.

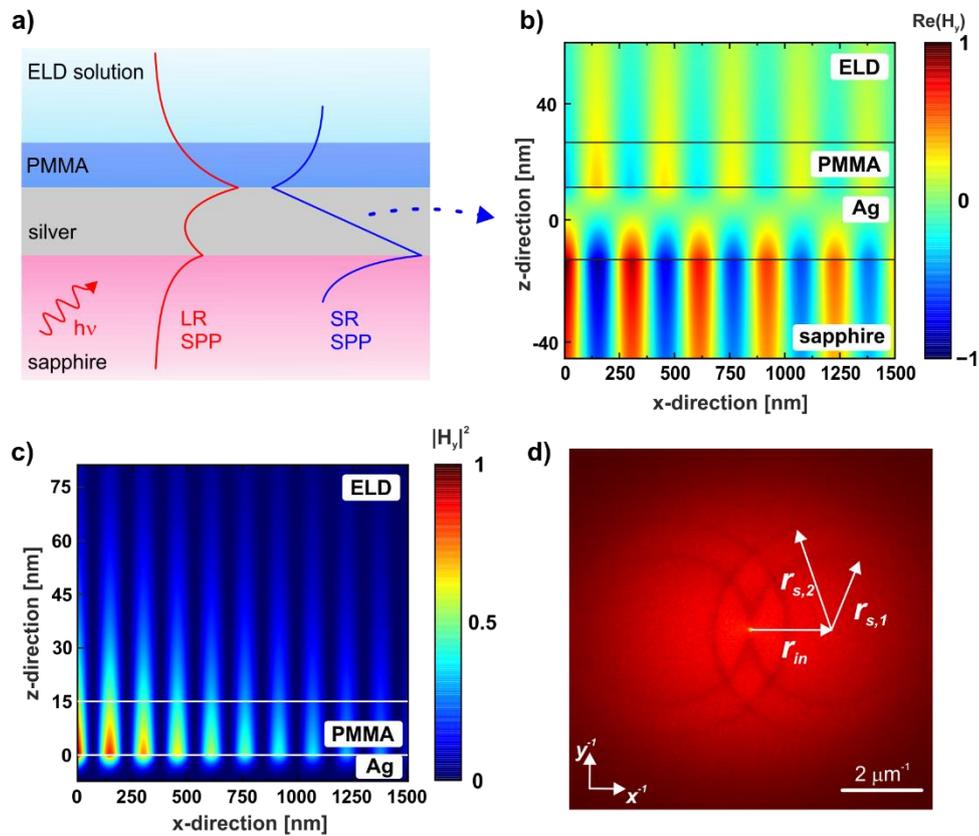

Figure 2: Metasurfaces deposited on thin silver films ($t_m = 25$ nm). a) The layer stack used for the experiments including qualitative field profiles of LR-SPPs (red) and SR-SPPs (blue); b) 2D field distribution of the SR-SPP mode simulated using RCWA; c) The corresponding 2D field intensity distribution; d) A typical FTEM of the metasurfaces fabricated on thin silver films exhibiting four structure rings.

The shift of the ring centers $r_{in}$ representing the exciting plane wave is identical for both pairs of rings, while the ring radii representing the scattered spherical wave differ. We therefore conclude that in the Kretschmann-Raether configuration only the LR-SPP can be excited directly. However, the metasurface provides the required momentum to excite SR-SPPs as well so it mediates coupling between LR-SPPs and SR-SPPs. Thus, it is experimentally proved, that features corresponding to the high index SR-SPP mode ($n_{SR\ SPP} = 2.1$ @$t_m = 25$ nm) can be incorporated into the resulting metasurface.

The variation of $t_m$ further allows for characterizing the dispersion of the hybrid modes with respect to the thickness of the silver film. We have fabricated silver films with thicknesses ranging from $t_m = 20$ nm to $t_m = 56$ nm. In **Figure 3a**, FTEMs of metasurfaces deposited on silver films of three different thicknesses are shown. It becomes visible that the outer structure ring diameter grows with decreasing $t_m$, while the radius of the inner structure rings stays largely constant in the case of (i) and (ii). This trend is also valid for larger $t_m$ (see Figure 3b), until hybridization decreases and the hybrid modes segue into the SPP$_a$ and SPP$_b$ whose mode indices are constant over $t_m$. In our experiments, we could observe hybridization effects up to a silver film thickness of $t_m = 46$ nm. For $t_m = 20$ nm, the effective index of the SR-SPP is as high as 2.3 and according to the simulation the corresponding penetration depth into the ELD solution is only 28 nm. In Figure 3b, a comparison between the effective indices of the plasmonic modes in simulation and experiment shows that the experimental results fit well to the results of the RCWA. Especially the good fit considering the effective indices of the SR-SPPs at thin silver thicknesses suggest that the mode propagation is largely undisturbed by the metasurface.

Only for $t_m = 20$ nm, we have observed a notable deviation of the simulated effective index of the LR-SPP and the measurement (see Figure 3b). This deviation was reproducible and robust against variations in the surface roughness of the silver. We have confirmed this robustness by comparing silver films deposited using different methods resulting in a surface roughness of 1.5 nm rms (thermal evaporation) and 1 nm rms (sputtering). The surface roughness has been characterized by atomic force microscopy.

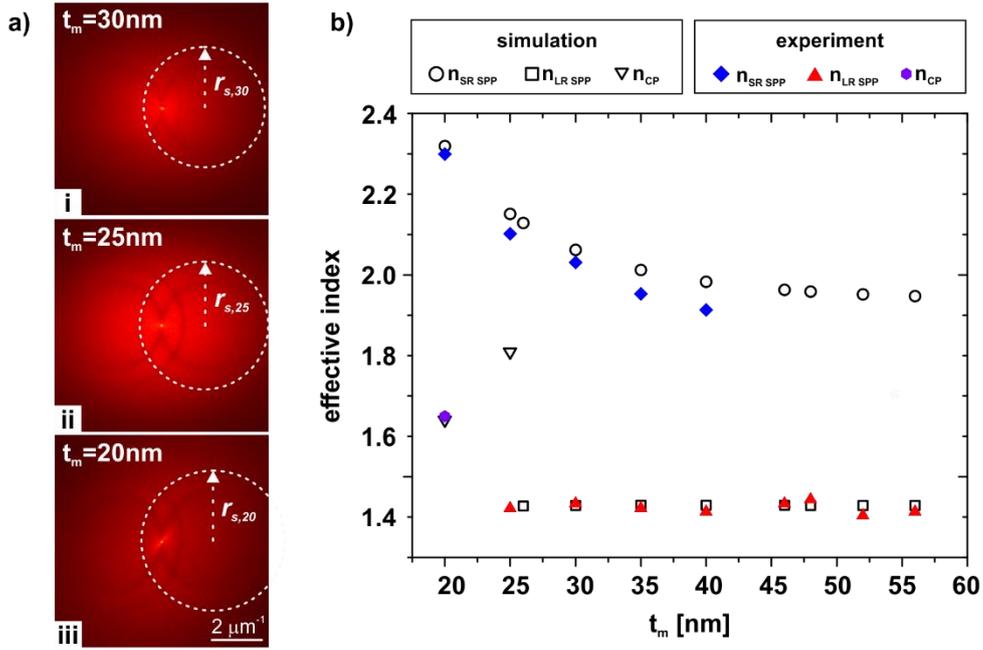

*Figure 3: Dispersion of the hybrid plasmonic modes with $t_m$. a) FTEMs of metasurfaces fabricated on silver films of 30 nm (i), 25 nm (ii), and 20 nm (iii) thickness; b) Measured and simulated effective indices for LR- and SR-SPPs.*

In the simulation results, we have found a counter-propagating mode (propagating in –x direction) showing an antisymmetric field profile (see Supporting Figure S1) and an effective index of $n_{cp} = 1.64$ (experiment: $n_{cp} = 1.65$). Although this finding suggests that multiple modes, which are accessible by the additional momentum of the metasurface are involved in the process, a detailed discussion of this mode is beyond the scope of this paper. In the simulation, the expected LR-SPP could not be found for silver films with thicknesses $t_m < 26$ nm. The fact that we have observed features originating from the LR-SPP mode for $t_m = 25$ nm in experiment (see Figure 3b) can be attributed to the limited measurement accuracy of the profilometer, surface roughness, and permittivity fluctuations, which are not taken into account in the simulation. However, in order to confirm the statement that multiple modes at largely different mode indices and different propagation directions can be involved in the process, we have performed additional experiments illuminating the samples with two lasers at different wavelengths, simultaneously. **Figure 4a** illustrates the corresponding experimental setup. Here, we use a spherical plano-convex sapphire lens instead of the half-cylindrical prism to utilize plasmonic modes, which propagate approximately perpendicular to each other ($\phi \approx 90°$) within the x-y plane. The laser wavelengths are 532 nm (propagating in x-direction) and 660 nm (propagating in y-direction).

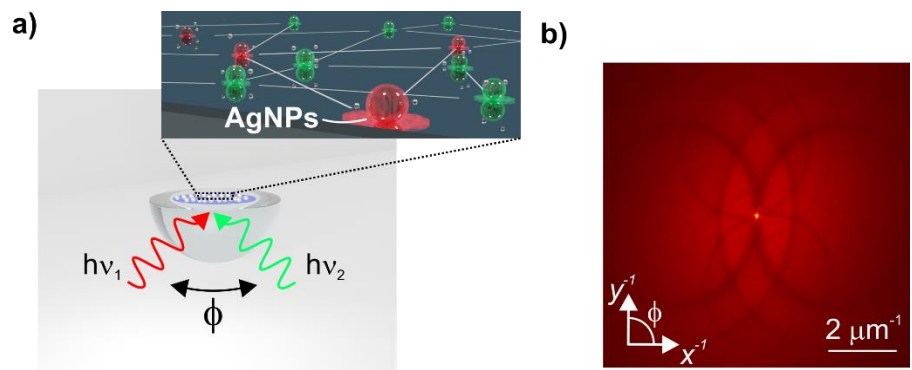

*Figure 4: Disorder-engineered metasurfaces fabricated on thin silver films using two excitation lasers simultaneously. a) A visualization of the experimental setup including the principle of the particle formation under illumination with different light sources (inset); b) A typical FTEM of the resulting metasurface showing eight structure rings corresponding to the four hybrid plasmonic modes involved in the process excited by two lasers at different wavelengths.*

Considering the FTEM of a metasurface fabricated using two lasers of different wavelength (see Figure 4b), it becomes clear that also multiple modes with largely different mode indices and different propagation directions in the sample plane can be utilized to incorporate features into the metasurface. In the scope of application, this is of particular interest, because it enables to exploit the functionality of various modes, each having a specific mode profile, penetration depth, effective index, and even frequency at the same spot simultaneously.

## 3. Conclusions

The utilization of short range surface plasmon polaritons in planar silver films has been shown by the plasmon-mediated growth of plasmonic metasurfaces controlled by light. The presented approach offers a facile way to utilize SR-SPPs by using bottom-up fabrication methods of an optical metasurface, while common top-down fabrication methods are often cost-intensive and technologically challenging.

The metasurface provides additional momentum, which enables to incorporate SR-SPP modes in the formation process of the metasurface. We recognize this involvement of SR-SPPs in the metasurface growth, because features of these modes (in the form of structure rings) appear in the FTEMs. We find mode indices simulated for planar structures without the metasurface that are in very good agreement with the experiment. This confirms that the metasurface does not significantly disturb the propagation of the present plasmon modes.

The functionality of the metasurface originates from its selective scattering properties that are highly sensitive to a change in the dielectric environment inside the mode volumes of the LR- and SR-SPP modes. Despite high sensitivities, SR-SPPs usually have limited sensing performance due to comparatively low $Q$-factors (broad resonance width), owed to short propagation lengths and thus strong localization. However, the approach presented here, enables to observe very sharp resonances, even if the modes involved have short propagation lengths. The sharpness arises due to the delocalized nature of the system response under coherent illumination on large areas (in the range of several tens of mm$^2$).[17] The results presented in this work thus represent a first step towards the facile utilization of SR-SPPs in planar optical sensing devices with extraordinary performance in detecting surface binding events in the future.

## 4. Experimental Section

*Sample preparation*: Sapphire substrates of thickness $t_S = 500$ µm are used for the experiments. First, an ultra-thin chromium adhesion layer ($t_{Cr} = 1.5$ nm) is deposited by sputtering. Then, a silver film is deposited onto the chromium film by physical vapor deposition (thermal evaporation) and a poly(methyl methacrylate) (PMMA) buffer layer ($t_{PMMA} = 15$ nm) is spin-coated on top of the silver enabling the growth of particles instead of film growth. For the experiments presented in Figure 3, the thickness of the silver films are varied between $t_m = 20$ nm and $t_m = 56$ nm. The film thicknesses are measured using a profilometer. Together with an index matching liquid, the sample is then mounted on a semi-cylindrical sapphire prism. The excitation of surface plasmon polaritons propagating along the silver/PMMA interface is realized by prism coupling (Kretschman-Raether configuration) of a p-polarized laser beam ($\lambda_0 = 660$ nm). We find the maximum excitation efficiency of SPPs (minimum in the ATR characteristic) under an incident angle of $\Theta_{in} = 54°$. The AgNPs are then grown from solution using the electroless deposition (ELD) technique. A detailed description of the preparation of the ELD solution for the plasmon-mediated growth of AgNPs can be found in an earlier work.[23] To start the particle growth, the ELD solution is drop cast onto the sample at the position of maximum optical power density ($I_{max} \approx 0.2$ W/cm²). After around 4 minutes of deposition (depending on $t_m$), the laser is switched off, and the ELD solution is removed from the surface. In the last step, AgNPs that have been randomly grown are removed by an ultrasonic treatment in a DI-water bath, enabled by their lower surface adhesion.[24]

*Scanning electron microscopy:* The SEM investigations were conducted using a Philips XL30S FEG system equipped with a field emission cathode.

**5. Acknowledgements**

This project has received funding from the European Research Council (ERC) under the European Union's Horizon 2020 research and innovation program (Grant Agreement No. 637367) and from the German Federal Ministry of Education and Research (Photonics Research Germany funding program, Contract No. 13N15390)

**References**

[1]   S. A. Maier, *Plasmonics: Fundamentals and Applications*, Springer, Berlin **2007**.

[2]   H. Raether, *Surface Plasmons on Smooth and Rough Surfaces and on Gratings*, Springer Berlin Heidelberg, Berlin **1988**.

[3]   D. K. Gramotnev, S. I. Bozhevolnyi, *Nat. Photonics* **2010**, *4*, 83.

[4]   S. Nie, S. R. Emory, *Science* **1997**, *275*, 1102 LP.

[5]   M. Moskovits, *J. Raman Spectrosc.* **2005**, *36*, 485.

[6]   J. Shi, Q. Guo, Z. Shi, S. Zhang, H. Xu, *Appl. Phys. Lett.* **2021**, *119*, 130501.

[7]   Y. Xu, P. Bai, X. Zhou, Y. Akimov, C. E. Png, L.-K. Ang, W. Knoll, L. Wu, *Adv. Opt. Mater.* **2019**, *7*, 1801433.

[8]   J. J. Burke, G. I. Stegeman, T. Tamir, *Phys. Rev. B* **1986**, *33*, 5186.

[9]   B. Fan, F. Liu, Y. Li, Y. Huang, Y. Miura, D. Ohnishi, *Appl. Phys. Lett.* **2012**, *100*, 111108.

[10]  A. Yanai, U. Levy, *Opt. Express* **2009**, *17*, 14270.

[11]  B. Frank, P. Kahl, D. Podbiel, G. Spektor, M. Orenstein, L. Fu, T. Weiss, M. H. Hoegen, T. J. Davis, F.-J. M. zu Heringdorf, H. Giessen, *Sci. Adv.* **2017**, *3*, e1700721.

[12]  A. Tuniz, M. Chemnitz, J. Dellith, S. Weidlich, M. A. Schmidt, *Nano Lett.* **2017**, *17*, 631.

[13]  M. Meudt, C. Bogiadzi, K. Wrobel, P. Görrn, *Adv. Opt. Mater.* **2020**, *8*, 2000898.

[14]  P. Berini, *Adv. Opt. Photon.* **2009**, *1*, 484.

[15]  J. A. Dionne, L. A. Sweatlock, H. A. Atwater, A. Polman, *Phys. Rev. B* **2005**, *72*, 75405.

[16]  J. Sukham, O. Takayama, A. V Lavrinenko, R. Malureanu, *ACS Appl. Mater. Interfaces* **2017**, *9*, 25049.

[17]  I. Shutsko, M. Buchmüller, M. Meudt, P. Görrn, *Adv. Opt. Mater.* **2022**, *10*, 2102783.

[18]  Y. Saito, J. J. Wang, D. A. Smith, D. N. Batchelder, *Langmuir* **2002**, *18*, 2959.

[19]  I. Shutsko, M. Buchmüller, M. Meudt, P. Görrn, *Adv. Mater. Technol.* **2022**, *7*, 2200086.

[20]  A. Cabarcos, C. Paz, R. Pérez-Orozco, J. Vence, *Powder Technol.* **2022**, *401*, 117275.

[21]  T. Itoh, N. Yamauchi, *Appl. Surf. Sci.* **2007**, *253*, 6196.

[22]  A. Polywka, C. Tückmantel, P. Görrn, *Sci. Rep.* **2017**, *7*, 45144.

[23]  I. Shutsko, C. M. Böttge, J. von Bargen, A. Henkel, M. Meudt, P. Görrn, *Nanophotonics* **2019**, *8*, 1457.

[24]  A. Polywka, A. Vereshchaeva, T. Riedl, P. Görrn, *Part. Part. Syst. Charact.* **2014**, *31*, 342.

# Supporting Information

**Harnessing Short-Range Surface Plasmons in Planar Silver Films**

**via Disorder-Engineered Metasurfaces**

*Maximilian Buchmüller, Ivan Shutsko, Sven Oliver Schumacher, and Patrick Görrn*

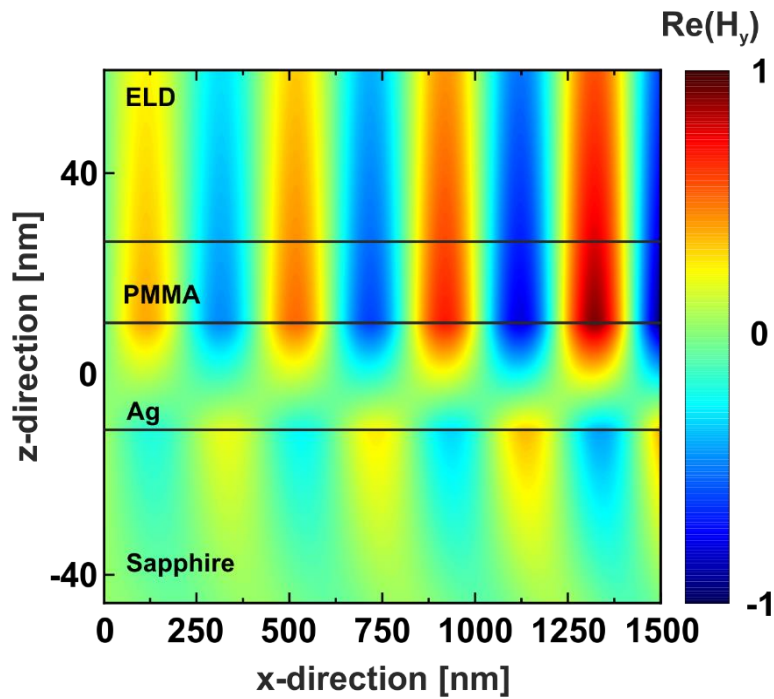

*Figure S1: 2D field distribution of the counter-propagating mode at an effective index of 1.64 for $t_m = 20$ nm.*

The field distribution displayed in Figure S1 is obtained using RCWA of the layer stack shown in the manuscript at a silver film thickness of $t_m = 20$ nm. The simulated effective index of the mode is $n_{cp} = 1.64$ and exhibits an antisymmetric field profile. Compared to the mode profile shown in Figure 2b of the manuscript, here the mode propagates in negative x-direction. In contrast to other anti-symmetric modes found in the simulation, here the mode is predominantly bound to the Ag/PMMA interface. Thus, the field extends far into the ELD solution and can thus strongly interact with the metasurface.